# Phylogenetic tree calculations using the Grid with DAG job

# Cálculos de árboles filogenéticos empleando la metodología DAG en un Grid

Raúl Isea*[1], Juan L. Chaves[2], Fernando Blanco[3], Rafael Mayo[3]

[1] Fundación Instituto de Estudios Avanzados - IDEA, Valle de Sartenejas, Baruta 1080, Venezuela.
[2] Parque Tecnológico Mérida, Av.4 Edif. Masini, Mérida 5101, Venezuela.
[3] CIEMAT, Avda. Complutense, 22, 28040 Madrid, España

(*) Corresponding author:

Raúl Isea. Instituto de Estudios Avanzados IDEA. Valle de Sartenejas, Hoyo de la Puerta, Baruta, Venezuela. Email: risea@idea.gob.ve

## Abstract

The goal of the work is to implement molecular phylogenetic calculations using the Grid paradigm by means of the MrBayes software using Directed Acyclic Graphs (DAG) jobs. In this method, a set of jobs depends on the input or the output of other jobs. Once the runs have been successfully done, all the results can be collected by a specific Perl script inside the defined DAG job. For testing this methodology, we calculate the evolution of papillomavirus with 121 sequences.

*Key words*: Directed Acyclic Graphs; MrBayes; Grid; Phylogenetic.

# Introduction

One of the more exciting challenges that these days are emerging in the computational biology is to determine the evolution history of the different species. One method for determination of the relationship among the species is Phylogenetics (Woese, 1998; Maher, 2002). As an example, we can mention the work from Korber *et al*. (2000), where the evolution history of the AIDS disease was determined. In this work, the authors deduced that AIDS doesn't come from a contaminated sample of a polio vaccine in Africa, but had its origin many years earlier (Hooper *et al*, 2000). This last result was obtained in a 512 Origin 2000 CPUs cluster called Nirvana sited at Los Alamos Advanced Computing Laboratory by running a modified version of fastDNAml. This program was improved by adding parallel architectures and routines and a reversible base-substitution model (Korber *et al*, 2000).

According to this, we can conclude that there is a computational method, which is able to build the evolution history of any species. Unfortunately, this achievement was obtained with a high computational time cost and, for most of the cases, the scientific community is not even able to get access to the supercomputer infrastructures as used in Korber *et al*. (2000).

For this reason, a work in developing alternative computational efficient techniques for estimating the phylogenetics in a faster way has to be done. Among them, we can currently find the following methods: distance based, maximum parsimony, maximum likelihood or bayesian. In this work, the latter will only be used with the help of MrBayes software (available at

www.mrbayes.net). However, MrBayes is relatively new in the construction of phylogenetic trees as the reader can check in the pioneering work of Rannala et al. (1996). This methodology works with the Bayesian statistics previously proposed by Felsentein in 1968 as indicated by Huelsenbeck, a technique for maximizing the subsequent probability (Huelsenbeck *et al*, 2002). The reason for using this kind of approach is that it deals with higher computational speed methods so the possible values for the generated trees can all be taken into account any of them not ruling the others.

Up to now, there are several centres involved in Grid computing in the biomedical field, where some examples are EGEE (web site at http://www.eu-egee.org/) or EELA (web site at http://www.eu-eela.eu/) and their further phases. The latter aims to disseminate distributed computing technology and to share European and Latin American resources via communication networks built on both continents (mainly EELA). As an example, we can mention the alignment of nucleotide sequences by means of the BLAST tools (Hernández *et al*, 2007). However, there are no applications to our knowledge for calculating phylogeny in the Grid environment by the use of heterogeneous and distributed resources with the submission of a JDL job with Directed Acyclic Graph (DAG) dependencies (Sato, 2008). Therefore, this works intends to supply this lack since DAG has been probed as a powerful technique in Bioinformatics.

The diversity of papillomavirus (PV) types has been calculated (Villiers *et al*, 2004). Four years ago, in the context of a phylogeny, it was necessary to define the term “species”. As an example, species linked to human PV-2 are typically found in common skin warts. PV types that form a species linked to PV-16 are also found with a high percentage in cervical cancer and its precursor lesions, *i.e*. they are considered as “high risk” components (such as human PV-18, human PV-45 and so on). In this context, this work has been done with 121 sequences from different PVs based on the nucleotide sequence of the major capsid gene, the most conserved gene in PV genomes.

Finally, it is important to mention that this work is not focused on the selection of the sequences and model in the phylogenetic context, but on the development of a new method which could perform phylogenetic calculations in a more efficient way by means of the Grid technology.

**Methods**

The sequences used in this work were downloaded from GenBank database, and lately, all of them were grouped in a unique Multifasta file. The next step was to align these sequences by means of Clustal software (Thompson et al, 1997) and to store the output file in nexus format (called ha.nex), which is the one that MrBayes uses.

The DAG technology was performed in a little 4 nodes cluster where three out of them were used as workstations for doing three independent calculations. They were labelled as A, B and C. These three calculations or scripts correspond to the three independent calculations of the phylogeny (see figure 1). It means that we run three independent calculations identified as ha***i*.**run, where *i* runs from 1 to 3 and corresponds to A, B and C, respectively. These three input scripts are built in a way that all of them are independent from the rest and are started from a random generated tree; by doing this, iterations in the input files are avoided.

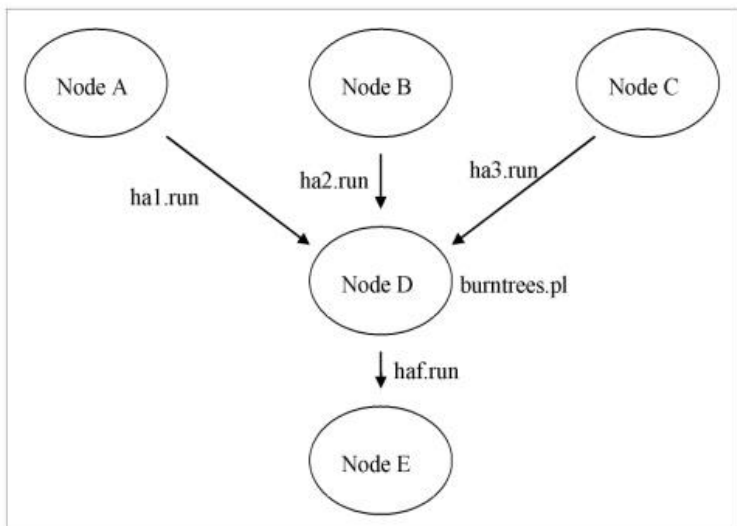

**Fig 1**. Schema of the DAG dependencies used in this work

The calculations starts in the nodes A, B and C and their corresponding results are then used as input for the node D, which is a central node of this cluster. We have only used three different nodes in order to validate the use of DAG in these phylogenetic Grid calculations, but of course this number can be increased (Fig. 1).

This input file in node D contains all the results from the previously generated executions and are stored in a concrete variable called OutputSandboxBase in our example. This variable can be stored and used in an improved GridFTP called Grid Security Infrastructure File Transfer Protocol (GSIFTP). Later, the Perl script developed by Johan Nylander called burntrees.pl [http://www.abc.se/~nylander] is able to manipulate in an independent way the phylogenetic trees deployed by MrBayes and put together all of them without loss of information.

In the last step (identified in Fig. 1 as node E), MrBayes infers a new calculation with a new script called haf.run where the final phylogenetic tree derived from the three previous ones is obtained. This final calculation contains implicitly in this way the three previous results performed in nodes A, B and C. This can be done because a consistency must be achieved; for this reason, the burning variable used in haf.run is exactly the same as used in the burntrees.pl script (*i*.e. 100).

At the same time, a similar calculation was performed in a cluster supporting MPI with the same number and type of processors that were used by MrBayes in the DAG method. The reason is again two-fold: first, to reproduce the same topology and validate the calculation performed on the Grid environment; and second, make a comparison between the two methods related to the consumed time.

**Results and Discussion**

The study of molecular phylogenies is worldwide the most used method for classifying the different types of papillomavirus (Villiers *et al*, 2004); even more, this method of studying phylogeny is the only one available to classify the diversity of PV types, and for this reason we checked our methodology in a phylogenetic calculation.

The final phylogenetic tree generated with the DAG Grid job can be seen in Figure 2, where the high-risk component is just in the same group as it has been identified in human PV-16, human PV-18 and so on. This fact agrees with the paper published by de Villiers where the neighbour-joining phylogenetic methodology was used (Villiers *et al*, 2004). Since our goal is to validate the Grid technology in this kind of calculations, a clear comparison can be made between the two results.

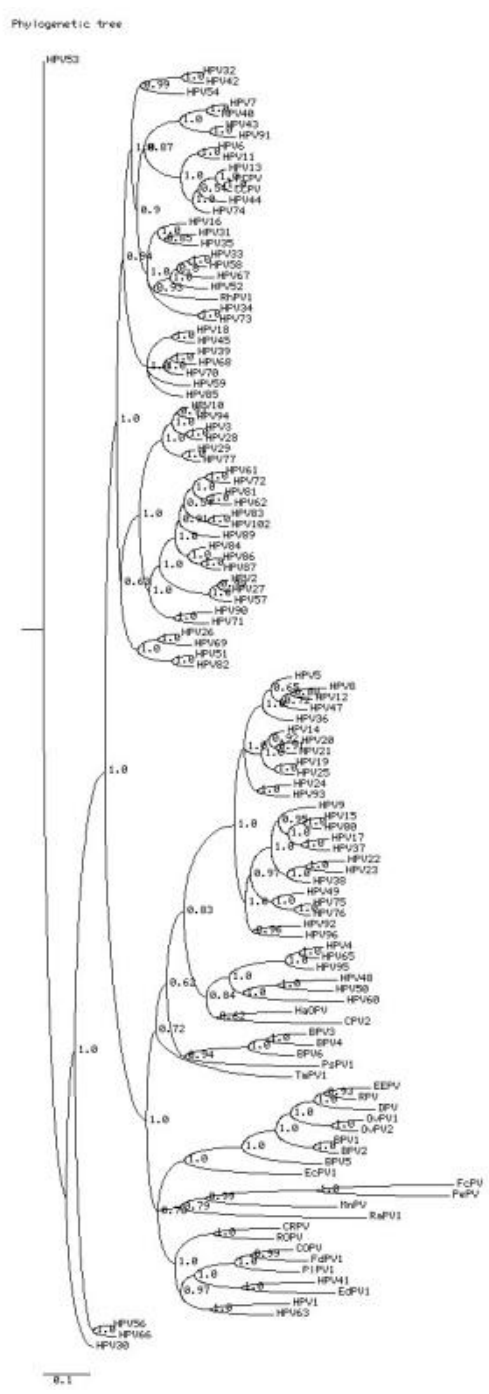

**Fig. 2**. Phylogenetic tree visualization of PVs with ATV software.[19]

As shown, both results are equivalent. By the use of the "comparetree" command, the outputs of the trees can be compared. Thus, using this command with our Grid DAG result and the previous published one, an agreement of 0.99718 (where 1.0 is identical) can be found. Thus, our results are consistent with the previous ones and, as a consequence, the use of the Grid can easily been inferred.

In our example, Intel Dual Xeon 2.3 GHz have been used and one node is able to calculate about 1 million steps in 5 hours. Since the Grid method is able to distribute the calculations to many Working Nodes at the same time, a high benefit of this technique is expected when the calculations will be extended to a greater number of inputs in the DAG dependencies. It is important to point out that only three nodes haven been used in our Grid job.

If we compare the time consumed by the Grid and the local cluster calculations, we find that the ratio of time used for achieving our phylogenetic result has been 1.21, *i.e.* the Grid is slower. The reason for this lays on the fact that some Perl scripts must be performed independently in order to collect the different calculations as if they were being executed in a single batch job. In this way, there is a master node which rules the whole process.

We must also take into account the drawback that for this work we were obliged to monitor the nodes to be sure that the program did not crash or stop with a stand-by. The longer time needed will be avoided in the future by means of the use of meta-scheduler such as Gridway (Huedo *et al*, 2005). This tool also allows the user to make advances in the load balancing or the assignment of nodes, so we shall be able to avoid the use of new scripts for such kind of tasks.

## Conclusions

This work demonstrates the efficient use of Grid Technologies for performing molecular phylogenetic calculations in many computational resources. For doing this, a DAG script has been developed. It takes advantage of the possibility of deploying distributed calculations in different allocated resources obtaining at the end a unique result.

The time used for achieving this result has also been only 1.21 times slower than the one used for the similar calculation in absence of the Grid environment.

Finally, initiatives such as EELA2 offering to the scientific community computational time in an easy way via a friendly interface for the researchers, allow them, to submit from their own computer high-demanding scientific calculations. The easy access is a key factor since most of the researchers are not familiar to the distributed computational techniques such as Grid.

## Acknowledgments

We are grateful to Johan Hoebeke who provided helpful comments on a previous draft of this paper. Authors thank in particular the support provided by the EELA Project (E-infrastructure shared between Europe and Latin America, http://www.eu-eela.org), contract n° 026409-6th Framework Programme for Research, Technological Development and Demonstration (FP6).